\documentclass[12pt,preprint]{aastex}

\def\mcol{\multicolumn}

\def\mult{$\times$}

\begin{document}
\title{Young stars and dust in AFGL437: 
NICMOS/HST polarimetric imaging of an outflow source}

\author{Casey A. Meakin, Dean C. Hines, and Roger I. Thompson}
\affil{Steward Observatory, University of Arizona, Tucson, AZ 85721}

\begin{abstract}
  We present near infrared broad band and polarimetric images of the compact star
  forming cluster AFGL437 obtained with the NICMOS instrument aboard HST. 
  Our high resolution images reveal a well collimated bipolar reflection nebulosity in 
  the cluster and allow us to identify WK34 as the illuminating source.   
  The scattered light in the bipolar nebulosity centered on this source is very 
  highly polarized (up to  79\%).  Such high levels of polarization implies a 
  distribution of dust grains lacking large grains, contrary to the usual
  dust models of dark clouds. We discuss the geometry of the dust distribution 
  giving rise to the bipolar reflection nebulosity and make mass estimates for the 
  underlying scattering material. We find that the most likely inclination of 
  the bipolar nebulosity, south lobe inclined towards Earth, is consistent with the inclination 
  of the large scale CO molecular outflow associated with the cluster, strengthening the 
  identification of WK34 as the source powering it.
\end{abstract}


\section{Introduction}

Polarimetric images of star forming regions have revealed nebulosities with
very high degrees of linear polarization which can only be accounted for by light 
scattering from small dust particles at nearly right 
angles (e.g., Werner, Capps \& Dinerstein 1983).  Many of these 
reflection nebulosities are associated with young optically invisible
protostars, and in some cases even near-infrared invisible protostars. The polarization 
signatures associated with these reflection nebulae offer valuable clues about the 
spatial distribution of scattering material and the relative location of the 
illuminating source(s), sometimes betraying the location of previously undetected and 
highly obscured objects (e.g., Weintraub \& Kastner 1996a). Polarization data can
also be used to constrain the composition and size distribution of the dust 
grains responsible for scattering (e.g., Pendleton, Tielens \& Werner 1990;
Kim, Martin \& Hendry 1994, KMH).

AFGL437 is a well studied compact cluster ($\sim$0.2 pc across) of young
stars and reflection nebulosity 2$\pm$0.5 kpc from the sun (Arquilla
\& Goldsmith 1984, and references therein). The youth of this cluster is inferred
from the presence of B stars and associated HII regions detected using
optical spectroscopy (Cohen \& Kuhi 1977). Three high luminosity members were 
found in an infrared survey (Wynn-Williams
et al. 1981), two of which are counterparts to the optical reflection
nebulosity studied by Cohen \& Kuhi. Other signs of active star formation include 
radio CO observations (Gomez et al. 1992)
of a broad, large scale (\( \sim  \) 1 pc) bipolar molecular
outflow coincident with the infrared cluster, and observations of 
water masers associated with cluster members (Torelles et al., 1992).

Several dozen deeply embedded, lower luminosity members of the cluster were
discovered through sensitive observations in the J, H and K bands by Weintraub
\& Kastner (1996b; hereafter WK1996). WK1996 also presented \( \sim 1'' \)
resolution ground based polarimetric images of the cluster in the J, H and K
bands which revealed polarization levels up to \( \sim  \) 50\% in the nebulosity
adjacent to one of the three high luminosity members, AFGL437N. The polarization
pseudo-vectors of the most highly polarized region showed a centrosymmetric
pattern centered on a before unseen intensity peak called WK34 by the authors.
Follow up 3.8\( \mu m \) images of the cluster with \( \sim 1'' \) resolution
revealed a nebulosity centered on WK34 which is extended along the molecular
CO outflow axis (Weintraub et al. 1996c).

We have performed follow up 1 to 2$\mu m$ polarimetric and broad band imaging of 
the reflection nebulosity associated with AFGL437 using HST/NICMOS.
Our observations reveal a very narrow bipolar nebula in the
field with polarization levels as high as 79\( \pm  \)3\% at 2 $\mu$m.
Following a presentation of the data we discuss the physical implications
of the observations and constraints placed upon the underlying mass
distribution.  This is followed by a summary of results and concluding
remarks.

\section{NICMOS Observations}

NICMOS polarimetric and broad band images of the central region 
of the AFGL437 IR cluster were obtained with 5 observations between Dec 30,
1997 and Feb 15, 1998 utilizing both the NIC1 and NIC2 cameras.
The log of observations, the cameras used for each filter and the measured PSF FWHM and
rms noise levels are presented in Table 1.
The 1$\mu$m and 2$\mu$m polarimetry  was obtained with the NIC1 and NIC2 cameras, respectively.  
Images of a blank patch of sky were observed through the F222M filter and the 2$\mu$m 
polarizers for background subtraction. The plate scales are $\sim$0.043\arcsec 
for camera 1 and $\sim$0.076\arcsec  for camera 2. The resulting field of view at 1$\mu m$ is
11.1\arcsec$\times$10.9\arcsec and the field of view for all other wavelengths is 
19.5\arcsec $\times$19.3\arcsec.

Our raw MULTIACCUM image data was calibrated using custom IDL procedures to
perform dark subtraction, cosmic ray removal, linearity correction and flat
fielding (Thompson et al. 1999). The field was imaged in a 7-position spiral-dither
pattern with a \( 0.27'' \) offset to sample over bad pixels and to average
over sub pixel sensitivity. The most contemporaneous dark frames and flat fields
maintained by the NICMOS Instrument Definition Team (IDT) team were used for the 
reduction. The photometric conversion from ADU s\( ^{-1} \) pix\( ^{-1} \) to 
Jy pix\( ^{-1} \) are given in Table 1 (M.Rieke 2000, private communication). The images 
were rectified to correct for the rectangular pixels projected onto the sky by 
the NICMOS detector array (Thompson et al. 1998). The rectification step
is essential for producing unskewed polarization position angles across the
field of view (Weintraub et al. 2000a; Hines, Schmidt \& Schneider 2000, hereafter
HSS2000).

\subsection{Polarimetry}

Our polarimetric data was reduced using the HSL algorithm as described in HSS2000.
The error analysis for NICMOS polarimetry
is expounded in great detail in HSS2000 as well as Sparks \& Axon (1999, hereafter SA1999).
These errors are discussed in the presentation of the data in the following
two sections. The foreground linear polarization expected
at near infrared wavelengths due to grain alignment in the intervening interstellar
medium is observed to be very low (\( \le  \) 4\%) for line of sight extinction
through dark clouds with \( A_{V}\le 20 \) (Goodman et al. 1995; see also Weintraub,
et al. 2000b for a discussion). Therefore any foreground polarization in
our data is expected to contribute (or cancel) only a small fraction of 
that which we have observed.

\subsubsection{2$\mu m$ Polarimetry}

\begin{figure}
  \plotone{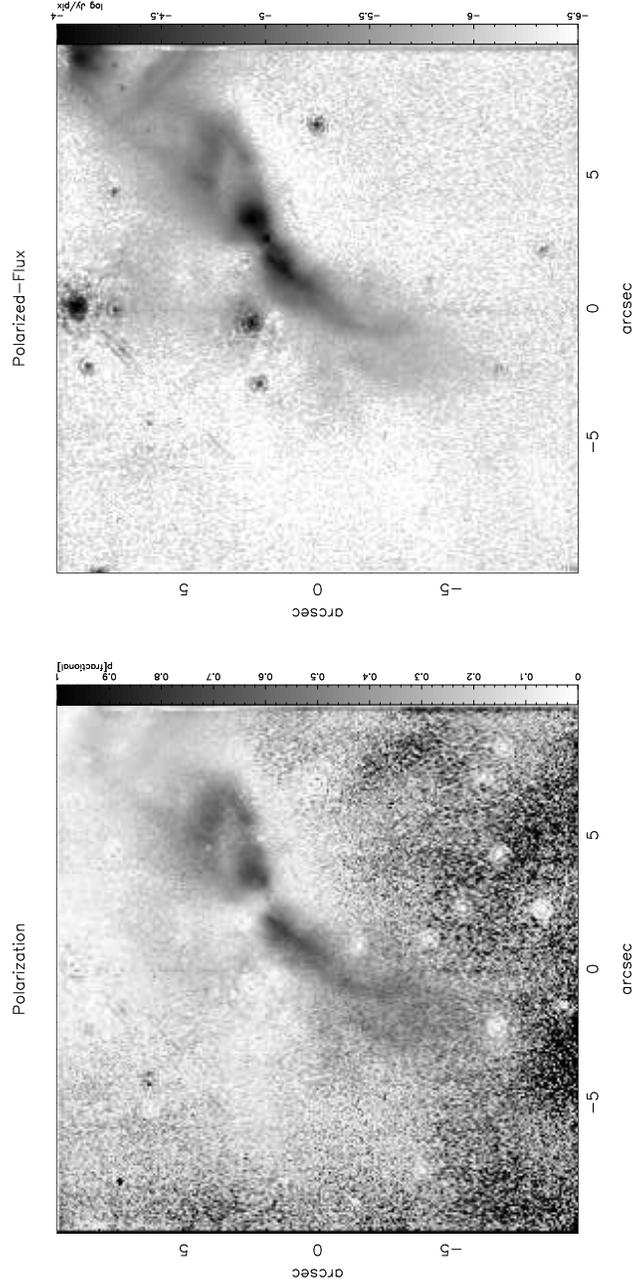}
  \caption[]{{\it Left:} Fractional polarization. {\it Right:} Polarized flux (total 
    intensity times fractional polarization).\label{f1}} 
\end{figure}

\begin{figure}
  \plotone{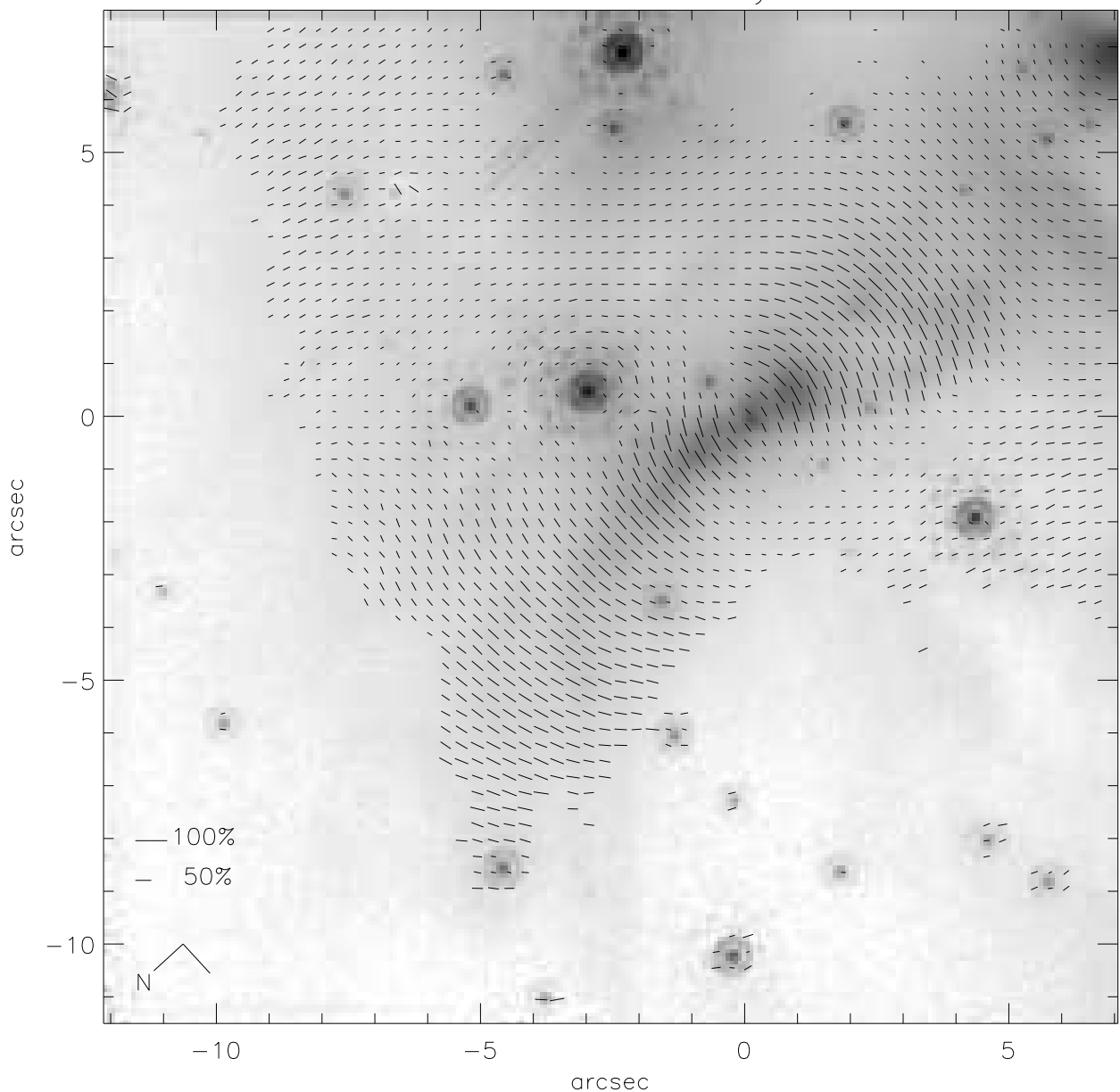}
  \caption[]{Polarization pseudo-vectors overlaid on logarithmic grayscale 
    of total intensity image. Vectors are plotted for $P>10\%$ and $S/N>10$. 
    The axes are plotted relative to the source WK34 which is located at the waist of 
    the prominent bipolar nebula. The  bright sources AFGL437N and W 
    are located at the approximate offset positions (-3.0,0.5) and (-2.4,7.0), 
    respectively, while the source AFGL437S resides just outside the image field a fraction of an arcsecond 
    to the south-east of the bright patch of nebulosity in the upper right hand corner of the 
    field.
    \label{f2}}
\end{figure}

\begin{figure}
  \plotone{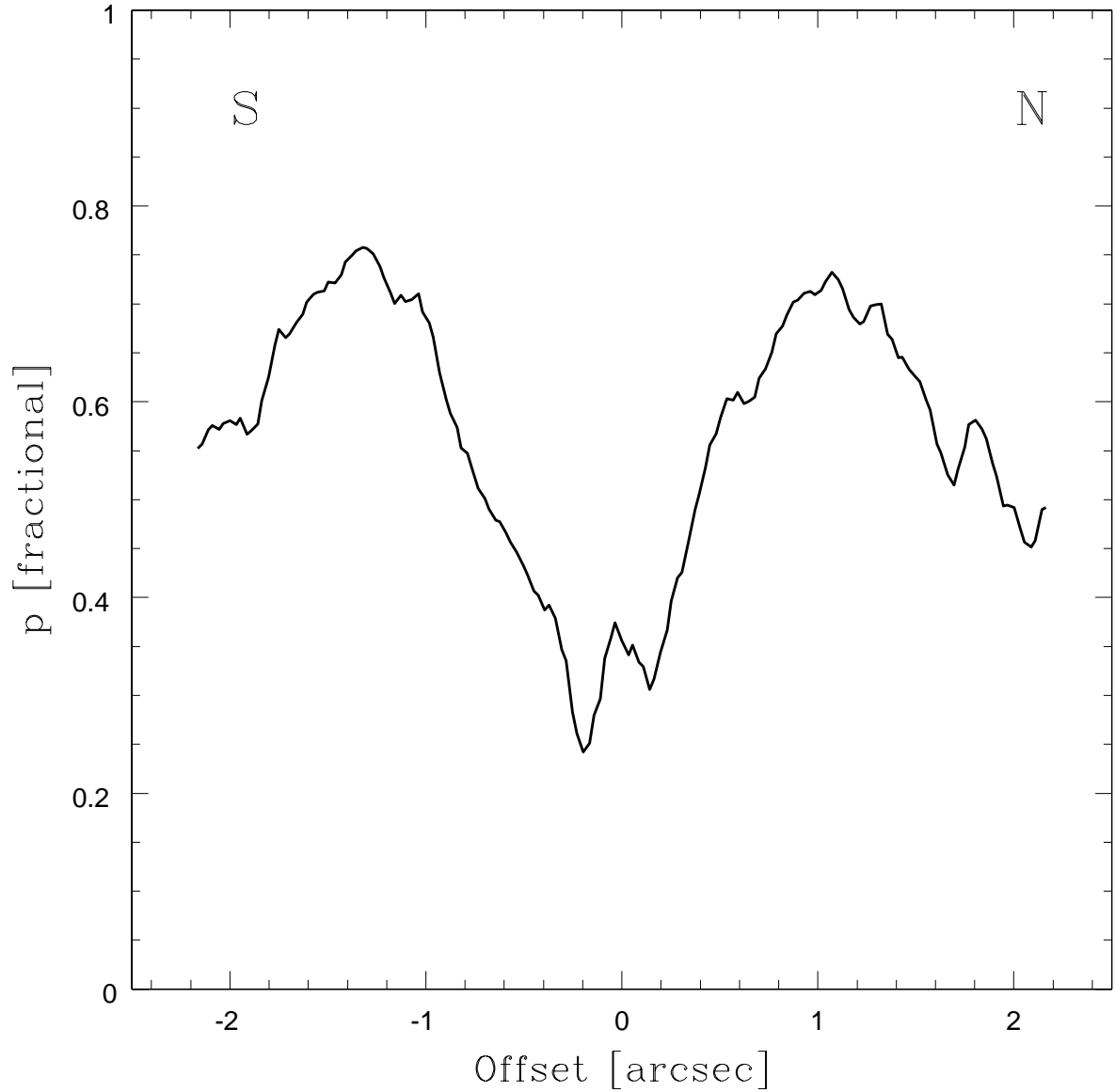}
  \caption[]{Fractional polarization for a slice along the bipolar nebula.  The slice is taken
    along line {\em AB} in Figure \ref{f6} with distance is measure relative to position {\em O}.
    The direction of the slice is indicate with the labels N and S indicating north and 
    south directions, respectively.  The polarization properties have been calculated after 
    rebinning the raw data into 2x2 pixel bins. 
    \label{f3}}
\end{figure}

We present the polarimetric imaging results in Figures \ref{f1} and \ref{f2}. In Figure
\ref{f1} we present gray scale images of the fractional polarization and the polarized
flux (fractional polarization times total intensity). In Figure \ref{f2} we present
the polarization pseudo-vectors overlaid on a gray scale image of the \( 2\mu m \)
total intensity image measured through the polarizers (see image captions for
details). The measured polarization levels in the field are remarkably high
and reach a maximum value of 79\% and 75\% in the north and south lobes, 
respectively, of the bipolar nebulosity which is located near the center of the field.
The polarization levels along the bipolar nebula in the field is shown in
Figure \ref{f3} (the location of the slice is indicated in Figure \ref{f6} by the 
line {\em AB} which is discussed below).  The polarization properties have been 
calculated after rebinning the raw data into 2x2 pixel bins.

The errors in the measured polarization levels and the polarization angle are
functions of both the intrinsic polarization levels, \( P \), and the signal
to noise (S/N) of the individual polarizer images. Signal to noise levels are
\( >10 \) throughout most of the field and \( >100 \) in the regions of the
bright bipolar nebulosity. From the error analysis in SA1999 and HSS2000 we find
for \( S/N \sim 10 \) and \( P>50\% \) an uncertainty in \( P \) 
of \( \sigma _{P}\sim 15\% \) and a dispersion in the position angle, 
\( \sigma _{\theta }\sim 10^{o} \). For a \( S/N>100 \) and  \( P>50\% \) 
with no flat field problems, \( \sigma _{P}<1.4\% \) and 
\( \sigma_{\theta }<0.5^{o} \). Flat field residuals add an error 
and a conservative estimate will be a factor of two increase with
\( \sigma _{P}\sim3\% \) and \( \sigma_{\theta }\sim1^{o} \).

The high degrees of polarization observed in our NICMOS images are not commonly
measured from the ground and are made possible by the high spatial resolution
afforded by NICMOS/HST. With a native NIC2 camera resolution of 0.076\arcsec/pixel
for our 2$\mu$m observations and a measured FWHM of 0.175\arcsec, the effective
resolution after binning the data into 2x2 pixel bins is $\sim$0.35\arcsec  (see Table 1).
Ground based observations of the AFGL437 field were made by WK1996 and the highest 
value of linear polarization that they measured was \( \sim 50\% \) consistent with the
lower spatial resolution (\( \sim 1'' \)) of their images.  Similar ground-vs-HST/NICMOS
``beam-depolarizing'' effects were seen in observations of the Egg Nebula (Sahai et al. 1998).

The imaged field consists of many point sources and extended emission. The bright
point source near the center of the field is AFGL437N and near the top of the
image is AFGL437W (see Figure \ref{f2}). The source between the lobes of the bipolar 
nebulosity is WK34.  The nebulosity in the field is composed of a few main components. 
Centered on WK34 is an elongated narrow bipolar nebulosity extending in the N-S direction. 
The northern end of this bipolar structure veers slightly towards the east. The southern 
end of this bipolar nebulosity is seen to end in an east-west \char`\"{}ridge\char`\"{} of
higher surface brightness emission just north of AFGL437S. Small patches of
nebulosity also surround AFGL437N, S and W.

In Figure \ref{f1} we see that the the bipolar nebulosity contains the most highly polarized
emission.  The bipolar nebulosity is also the most prominent object in the polarized 
flux image (where polarized flux is the total intensity image multiplied by the
image of fractional polarization). Most of the polarization pseudo-vectors, shown 
in Figure \ref{f2}, form a centrosymmetric pattern around the center of the bipolar 
nebula. These pseudo-vectors deviate from centrosymmetry and align in an E-W direction
within a few arcseconds of the nebular waist. The polarization 
levels also drop to nearly zero a few arcseconds to the east and west of the nebular center.

Less pronounced centrosymmetric patterns are apparent around AFGL437N and, to a lesser
degree, AFGL437S, which lies just to the south of the observed field. 
The polarization levels in these regions, $P<20\%$, are much lower than 
surrounding WK34. The polarization pseudo-vectors align along the 
\char`\"{}ridge\char`\"{} of high surface brightness in the image at the intersection 
of the centrosymmetric pattern around WK34 and AFGL437S. The \char`\"{}ridge\char`\"{} 
may result from an interaction between these two sources (see also section 2.2).

\subsubsection{1 $\mu m$ Polarimetry}

Our 1 micron polarimetric images suffer from a low signal to noise (S/N of a
few) resulting in large uncertainties in polarization and position angle. 
Therefore, we restrict our attention in this data set to the polarization
and photometry in a circular 0.5\( '' \) radius aperture centered on the northern
lobe (the location of the aperture is shown in Figure \ref{f6}). For comparison we 
repeat this aperture measurement on the 2 micron data.
The S/N in the aperture is above 100. At 1 \( \mu m \) we measure 
\( P=45\pm 3\% \) with \( \theta _{p}=24\pm 1^{o} \) and at 2\( \mu m \) 
we measure \( P=65\pm 3\% \) and \( \theta _{p}=28\pm 1^{o} \).
The aperture polarimetric and photometric data are summarized in Table 2.
The southern lobe is contaminated by a point
source which is most evident in the ratio images in the bottom row of
Figure \ref{f7}, particularly the F160W/F110W ratio image. This point source
contributes a large fraction of the \( 1\mu m \) flux in the region so 
we ignore the polarization properties of the south lobe.

\subsection{Broad Band Images}

\begin{figure}
  \plotone{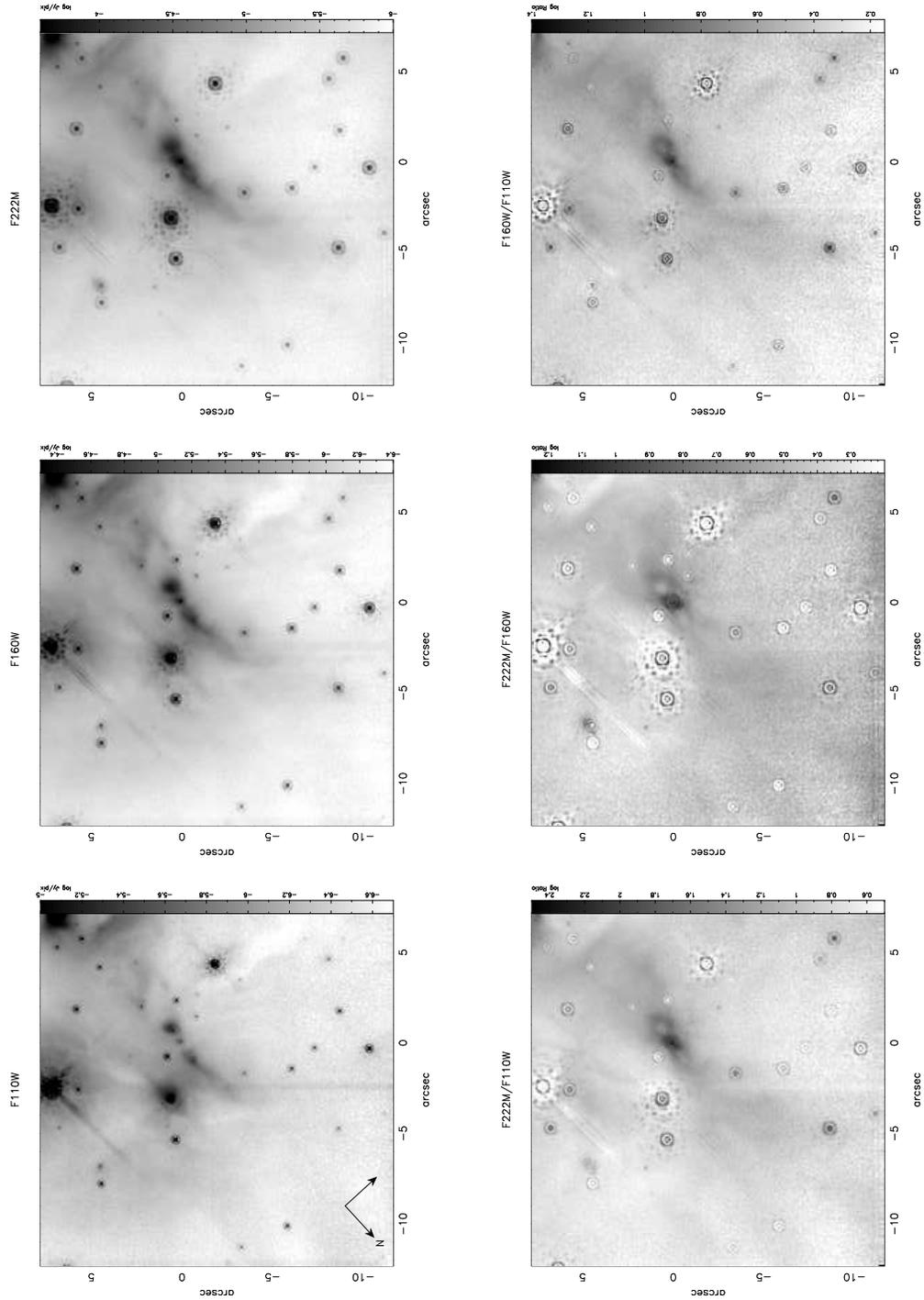}
  \caption[]{Broad band intensity and flux ratio images of entire field.\label{f4}}
\end{figure}

\begin{figure}
  \plotone{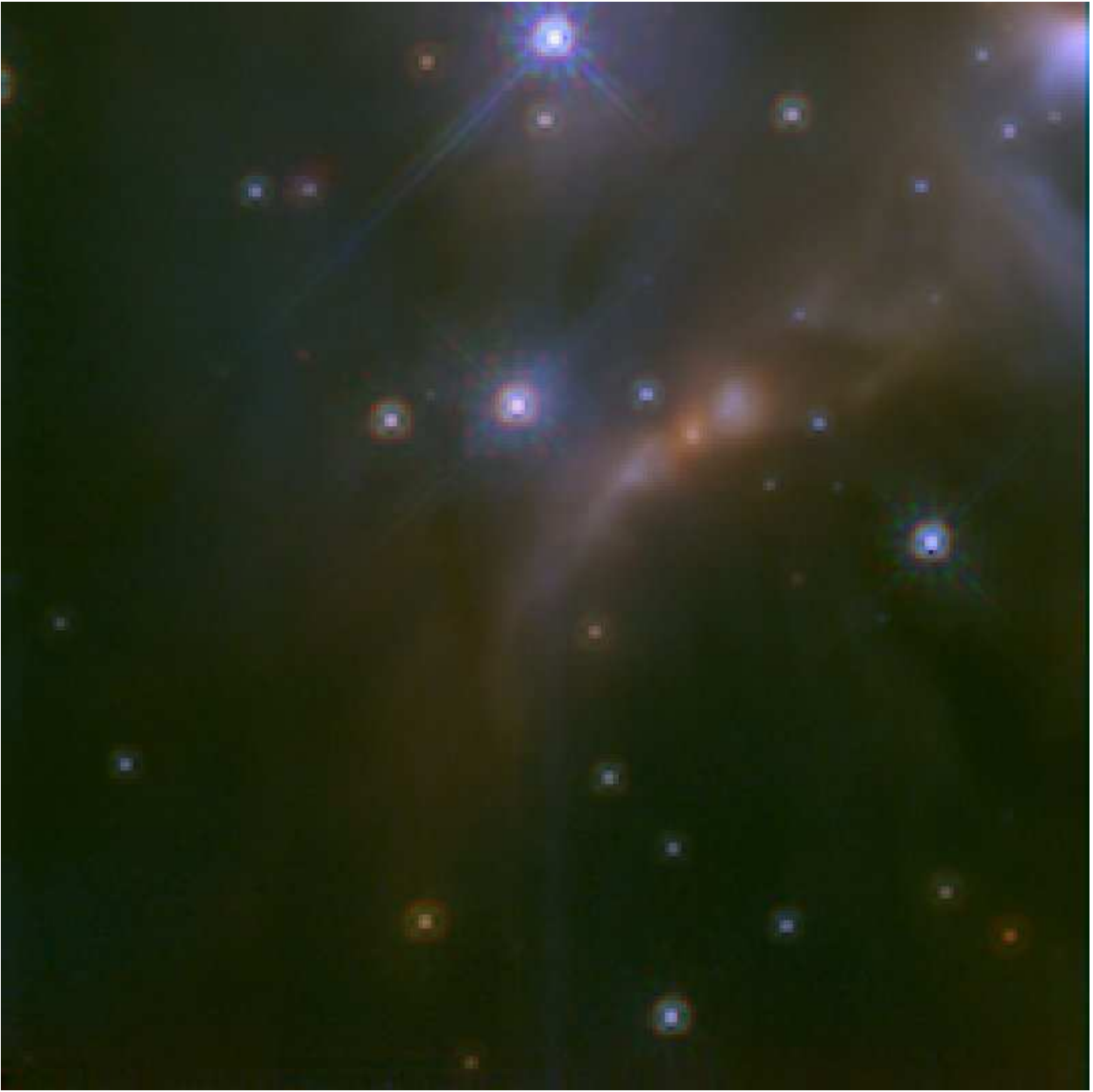}
\caption[]{Three color composite image of entire field. F222M(Red), F160W(Green), 
  and F110W(Blue) \label{f5}}
\end{figure}

In Figure \ref{f4} we present gray scale images of the intensity measured though the
F110W, F160W and F222M filters as well as the flux ratio images for the filter 
combinations F222M/F110W, F222M/F160W and F160W/F110W. In Figure \ref{f5} we 
present a three color composite image of the camera 2 data. 

The point sources and reflection nebulosity display a large range of flux ratios.
The nebulosity near the bright sources AFGL437N, S and W is relatively blue within 
the field and has an F222M/F110W flux ratio of \( \sim  \)10. Cohen and Kuhi (1977) 
found that the optical spectra of the extended emission in a $4''\times2.7''$ 
aperture centered on AFGL437N and S is consistent with slightly reddened 
(\( A_{V}\sim 6.5) \) B5 stars. We detect several faint, very red point sources 
in our field that were not detected in the WK1996 data of the same region.

\begin{figure}
  \plotone{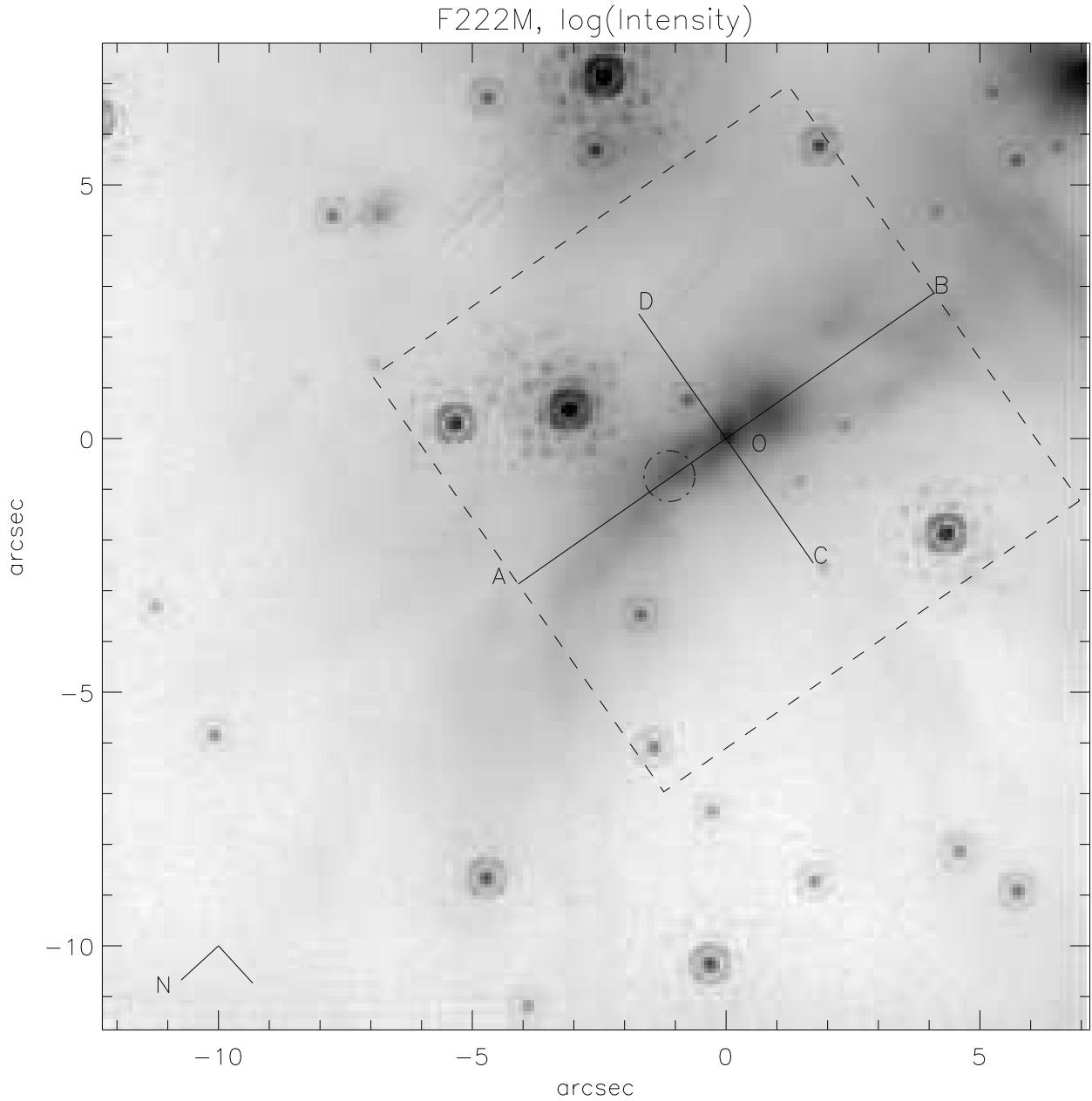}
\caption[]{Annotated F222M intensity image. The line {\em AB} is the axis fit to 
  the bipolar lobes and line CD is its perpendicular as discussed in text.  
  The dashed boxed shows the region presented in Figure 6. The circle indicates 
  the region where aperture polarimetry was performed (sec. 2.1.2). \label{f6}}
\end{figure}

\begin{figure}
  \plotone{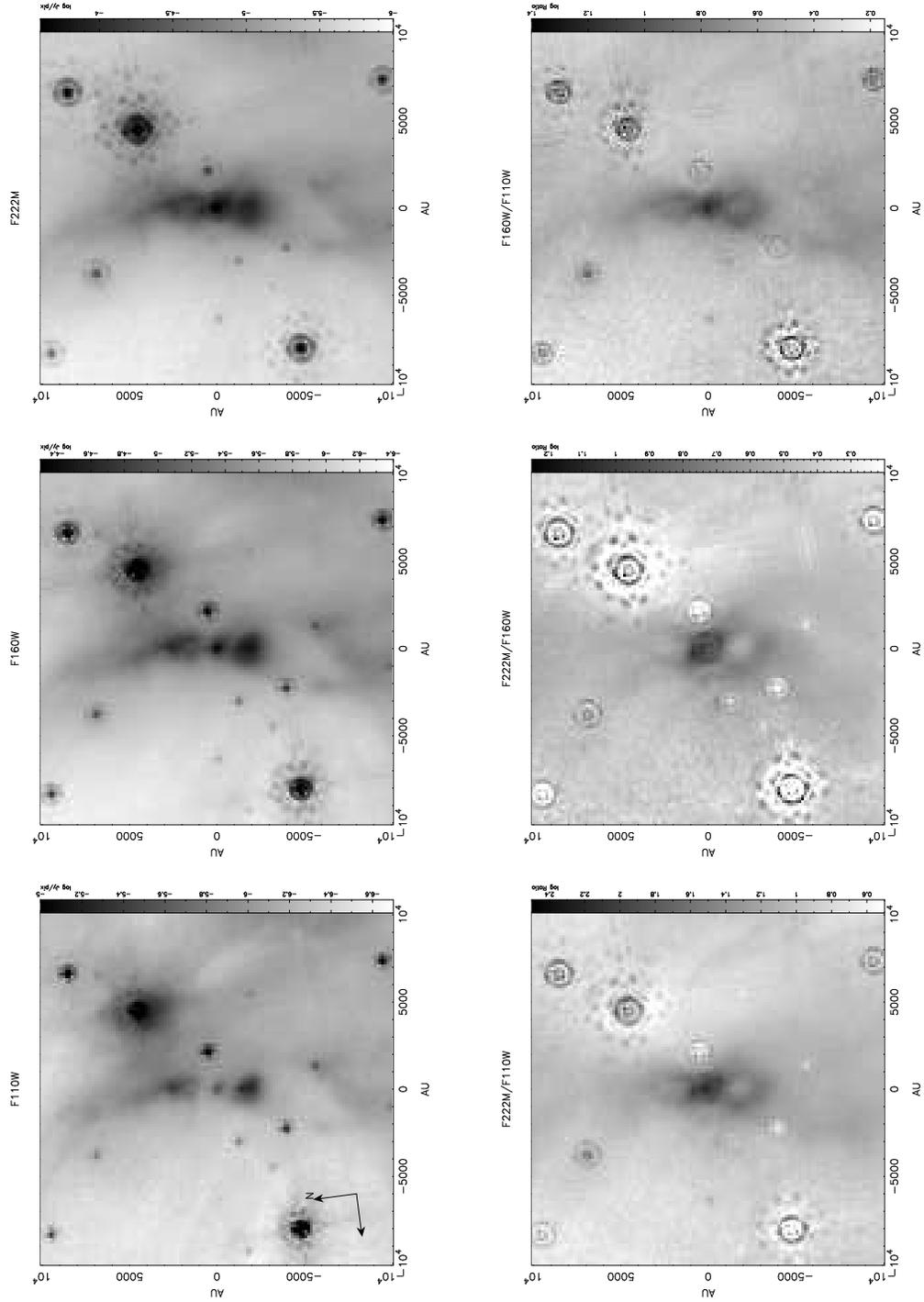}
  \caption[]{Broad band intensity and flux ratio images of WK34 detail. The location
    of this field is indicated by the dashed box in Figure \ref{f6} (see also compass
    for orientation).
    \label{f7}}
\end{figure}

The nebulosity associated with WK34 is extremely red with an F222M/F110W intensity
ratio as high as \( \sim  \)100 indicating either a high degree of extinction
towards this nebulosity or an illuminator with an intrinsically very red spectral
energy distribution. Details of the WK34 reflection nebulosity are presented
in Figure \ref{f7} with axes labeled in AU assuming a distance of 2 kpc. 
Figure \ref{f6} indicates the region and orientation of this closeup view
relative to the whole field. The data in Figure \ref{f7} has been rotated such that 
the principle axis of the bipolar nebulosity lies vertically. We fit this axis to the 
inner \( 10'' \) of the nebulosity where the axial symmetry is highest.  We constrained 
the axis to go through WK34 and the intensity peaks in the north and south lobes. 
The position angle we fit is 100$^{o}$ east through north. We over plot this axis on the 
F222M intensity image in Figure \ref{f6} as line {\em AB} and its perpendicular as 
line {\em CD}.


An abrupt change in color is seen across an east-west ``ridge'' of brighter emission 
in the southern end of the field (see the F160W image in Figure \ref{f4} at the offset 
position $\sim$(5.0,5.0) and compare with the color composite image in Figure \ref{f5}). 
The nebulosity north of the ``ridge'' is a relatively red feature in the field,
consistent with it being illuminated by WK34, while just south the nebulosity is much bluer 
consistent with the B5 spectrum of AFGL437S.  Patchy and filamentary dark features, 
possibly dust patches, are also apparent in the intensity and flux ratio images.

\section{Discussion}

\begin{figure}
  \plotone{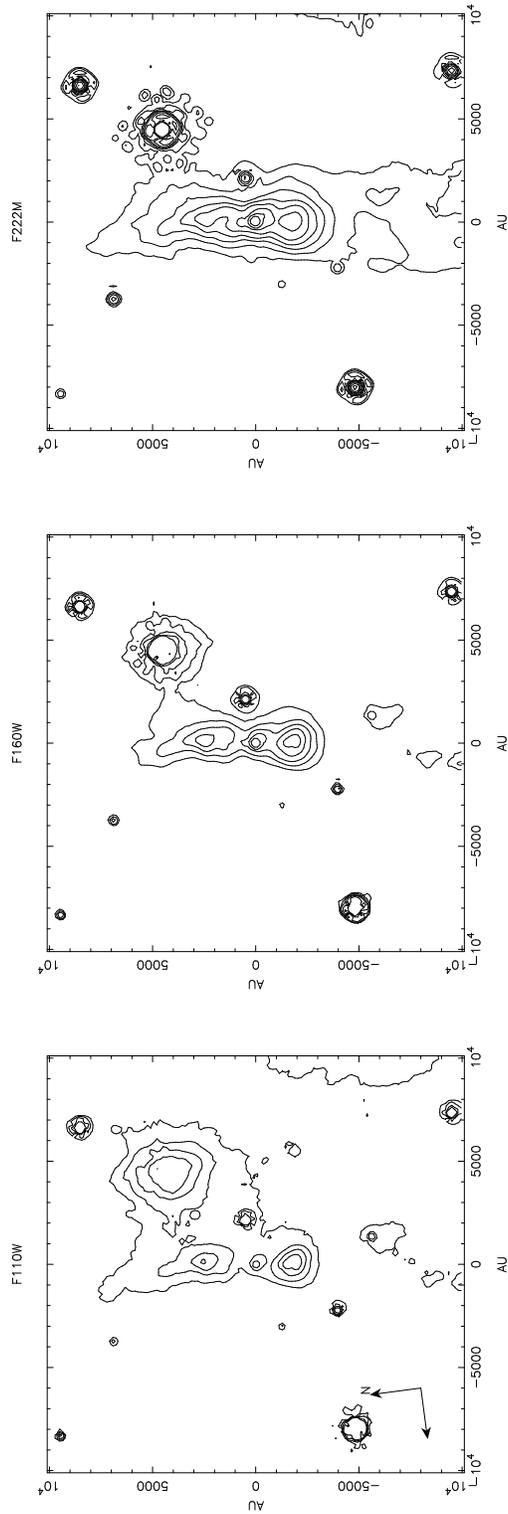}
  \caption[]{Contour plots of broad band intensity of WK34 detail for the same field
    as Figure \ref{f7}.  The contour 
    levels are separated by 0.5 magnitudes. For reference, the peak contour level in the
    southern lobe of the bipolar nebulosity is 3.589\mult10$^{-6}$, 3.02\mult10$^{-5}$, 
    and 6.76\mult10$^{-5}$ Jy/pix in the F110W, F160W and F222M image, respectively. \label{f8}}
\end{figure}

The most striking feature in the observed field is the bipolar reflection
nebulosity extending in the N-S direction near the center of the field. It
dominates the polarization and the polarized flux images (Figures \ref{f1} \& \ref{f2}).
In this section we discuss this prominent bipolar nebula 
and the source near its center, WK34. Complementary contour plots of the region centered
on the WK34 source (Figure \ref{f7}) are presented in Figure \ref{f8}. In the following 
subsections we discuss the nature of the illuminating source and the underlying mass 
distribution which gives rise to the reflection nebulosity.

\subsection{The Illuminating Source}

The polarization levels measured in the bipolar lobes are the highest
in the entire field with values in the north and south 
lobes of 79$\pm$3\% and 75$\pm$3\%, respectively. 
The centrosymmetric polarization pseudo-vectors which surround
the center of this nebulosity together with the very high percentage polarization
suggests that this nebulosity is being illuminated by a single source at its
center and that much of the illuminators light is singly scattered at near
right angles into our line of sight and have traversed optically thin
paths.

The geometry of single scattering allows us to calculate the position of the
illuminating source in the bipolar nebulosity by centroiding the normals to
the polarization pseudo vectors. This works because an unpolarized photon scattering
from a small particle (such as a dust grain) acquires linear partial polarization
in a direction perpendicular to the scattering plane (the plane containing the
paths of the incident and scattered photons).  We carried out a procedure to locate 
the illuminating source described in Weintraub et al. (2000a). We used
only those pixels with \( P>50\% \) and a \( S/N>50 \) for our centroiding
calculation. The position that we find for the illuminating source using this method
is within 2 pixels ($<0.15''$) of the centroid of the unresolved PSF 
at the equator of the bipolar object, i.e., WK34. This result strongly 
supports the original suggestion by WK1996 that this source is indeed 
illuminating the nebulosity. Furthermore, since we see an unresolved point source 
at this location it is very likely that we are imaging the source directly.

\subsection{The Nature of WK34}

The point like source at the waist of the bipolar nebula, WK34, is seen
in all of our broad band images. In both the F222M and F160W images we see
the characteristic NICMOS PSF superposed on a patch of nebulosity (Figure \ref{f7}).
The PSF is more difficult to discern in the F110W image but an intensity
peak is clearly apparent and its centroid is within a fraction of a pixel of
the centroid of the PSF in the F160W and F222M images. Photometry was performed in each filter 
by subtracting a scaled field star PSF from the same image.
The dominant source of uncertainty in this photometry
arises from the difficulty in gauging a good subtraction due to the 
patchy distribution of background nebulosity. The WK34 photometry 
with conservative error estimates is summarized in Table 3.

\par The emission from the WK34 source, presumably a young stellar object still in 
the outflow epoch, is likely to arise from multiple components including
a photosphere, an accretion disk, as well as heated ambient material
from which the star formed.  Light scattering by ambient material in the vicinity of
the star is also thought likely to affect the near infrared spectrum.  Model 
spectral energy distributions of low mass young stellar objects including these components of 
emission have been calculated by Adams, Lada \& Shu (1987). The WK34 photometry 
is compatible with these models which suggests that it is a low mass, low luminosity protostar.
Observations at longer wavelengths are needed for a robust measure of the bolometric
luminosity of this source.  Nevertheless, there is additional evidence that the WK34 source 
is a low luminosity protostar including: (i) that 
the bulk of the cluster luminosity (L=10$^{4}$L$_\odot$ at 2 kpc; Parmar et al.
1987) can be accounted for by the three early B ZAMS stars present in the cluster
and, (ii) that the WK34 source lacks an HII region (determined by contemporaneous 
\( P\alpha  \) NICMOS  observations to be presented in another paper) indicating 
the lack of a hard ionizing continuum usually associated with high luminosity sources.

\subsection{The Underlying Mass Distribution}

Many optical and near-infrared reflection nebulae associated 
with low mass protostars have been successfully modeled as
rotationally flattened protostellar envelopes with variously shaped 
evacuated cavities and an accretion disk (e.g. Whitney \& Hartmann, 1993; 
Kenyon et al. 1993; Wood et al. 2001).  
Associated model SEDs have been made for a wide range of wavelengths providing a 
theoretical underpinning to the multiwavelength studies of YSOs 
(Adams, Lada \& Shu, 1987).
We find, however, that similar models fail to reproduce the reflection nebula
associated with WK34.
The most striking features which distinguish the bipolar nebula in our field
are: (1) The appearance of very distinct, well separated lobes 
forming the bipolar nebula. (2) The distance between the intensity maximum
in each lobe is separated by a large distance ($\sim$ 5000 AU, for a
cluster distance of 2 kpc) relative to the overall nebular size. 
(3) The nebular lobes are much longer than 
they are broad.  Each of these characteristics requires a rotationally flattened envelope
model with parameters outside a reasonable range, indicating the inadequacy
of this model to explain our observations.
For example, modeling the nebula with a rotating 
collapsing cloud model such as that described by Tereby, Shu \& Cassen (1984)  
requires a centrifugal radius much larger than $\sim$2000 AU to 
explain the large separation between 
lobes, but then fails to produce the observed narrowness of the lobes.  After 
an exhaustive study of parameter space with a single scattering 
radiative transfer code we find that no reasonable combination of rotating envelope, 
evacuated cavities and accretion disk can explain the observations adequately.
The small effects of multiple scattering are likely to exacerbate the modeling
by contributing extra illumination at the nebular waist.

\par
\begin{figure}
  \plotone{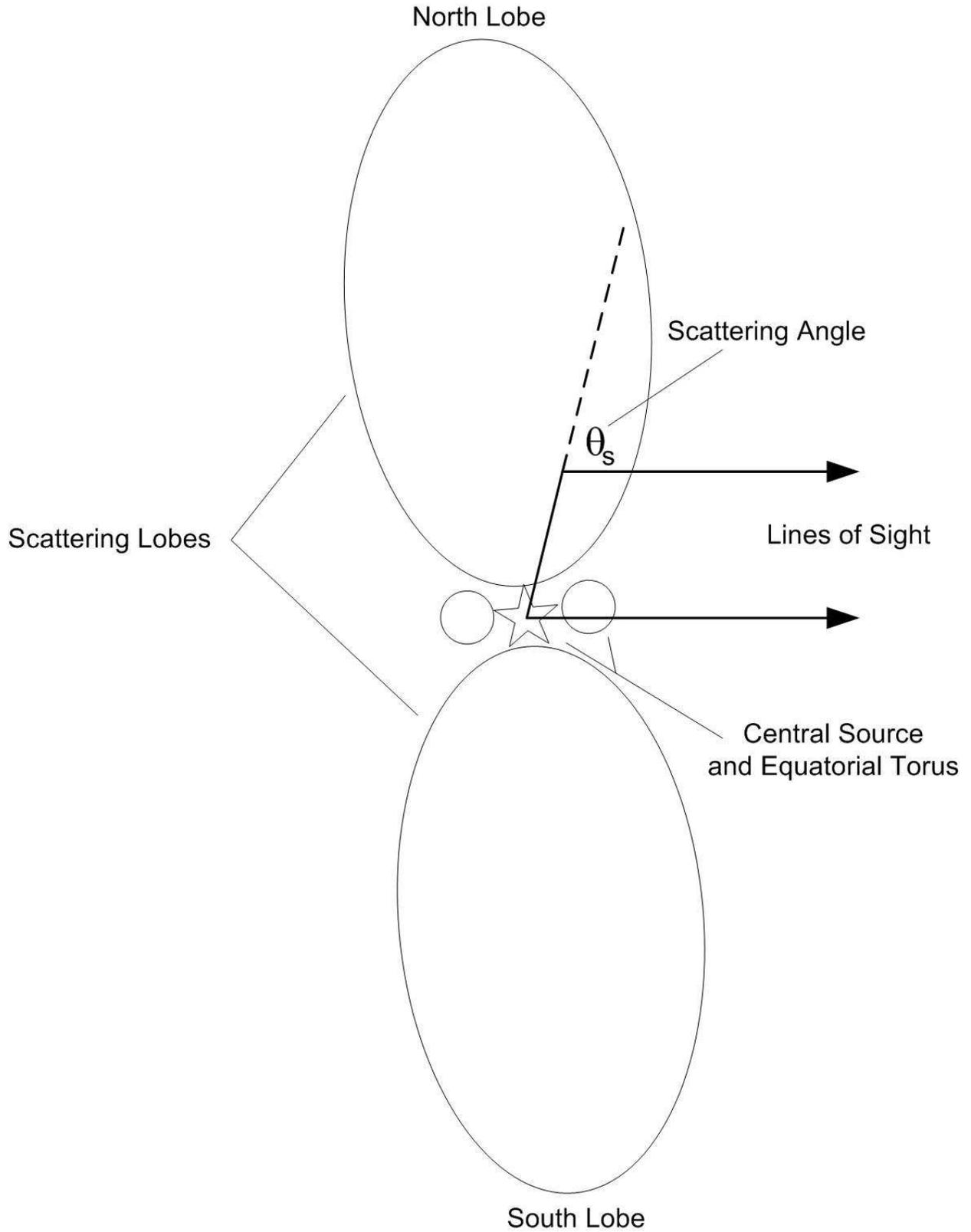}
  \caption[]{Schematic diagram depicting the geometry of the {\em outflow nebula} model proposed 
    to explain the prominent bipolar reflection nebula centered on the WK34 source.\label{f9}}
\end{figure}

 An alternative to the evacuated cavity model is one in which the bipolar reflection nebula 
is due to a large amount of outflow material ejected by the central source along the
nebular axis. The observed bipolarity of such an {\it outflow nebula} can then be further 
accentuated by the presence of an obscuring torus of material surrounding the central 
illuminating source, perhaps the remnant of the infalling parent envelope
(see schematic diagram in Figure \ref{f9}). This geometry is consistent with 
the pattern of polarization vectors
which deviate from centrosymmetry near the waist of the bipolar
nebulosity where they align along the equator of the nebula (see Figure \ref{f2}).
This type of polarization signature has been called a ``polarization 
disk'' and has been shown to be a phenomena of multiple scatterings
in the presence of an optically thick equatorial disk 
(Bastien \& Menard, 1988; Whitney \& Hartmann, 1993). This polarization 
signature arises on the observer side of an optically thick 
disk being illuminated from the bipolar lobes: we are seeing photons 
scattered {\em over} an optically thick disk into our line of sight.  The 
location of the ``null points'', or regions of low polarization on either
side of the ``polarization disk'', give us a rough estimate of
the size of the optically thick disk.  From Figure \ref{f2} we estimate
a disk radius of $\sim$1000 AU.  This pattern is common in the
bipolar nebulosity associated with both YSOs as well as protoplanetary 
nebulae such as the Egg Nebula (Sahai et al. 1998) and
OH231.8+4.2. The latter object, also known as the Rotten Egg Nebula because of its
sulfur content, is one of the few other astronomical objects with
measured infrared polarization levels as high as those presented here 
(Ageorges \& Walsh 2000; Meakin et al. 2003).

\par The high levels of observed polarization constrain the size of dust particles
responsible for scattering.  The grain model proposed to explain the extinction in molecular 
clouds by KMH is only capable of polarizing 2$\mu$m light to a maximum of $\sim$70\% 
(Weintraub, Goodman, \& Akeson, 2000b). This is due to a component of large particles in the 
grain size distribution of this model. It is likely that the particles in the reflection 
nebula observed here are composed mainly of particles much smaller than the observed 
wavelength.  The existence of small particles in the outflow lobes may be related to 
the outflow mechanism.  Dust grains may be selectively accelerated into the 
outflow material based on size, or they may be processed by shocks or photodisentegration 
processes which occur during the star formation process.

\subsubsection{Physical Parameter Estimates}

\par The relationship between the scattering optical depth through the lobe and
the density is $\tau_{los} = \bar{\rho_d}\Delta S\kappa_s$ where $\bar{\rho_d}$ is the
average lobe dust density, $\Delta S$ is the line of sight extent of the nebula, 
and $\kappa_s$ is the scattering opacity.  If we assume that the nebula is optically
thin, $\tau_{los}  < 1$, we have:
\begin{equation}
  \bar{\rho}_d < \frac{1}{\Delta S\kappa_s}.
\end{equation}

\par From Figure \ref{f8} we estimate a total lobe thickness $\Delta S \approx 2000$ AU, 
assuming that the nebula is roughly cylindrical and that the 2$\mu$m intensity image
traces its extent.  The mass scattering opacity for a spherical dust grain with
radius $a$ is,

\begin{equation}
  \kappa_s = \omega Q_{ext}\frac{\pi a^2}{\rho_i 4\pi a^3/3} = 
  \omega Q_{ext}\frac{3}{4a\rho_i},
\end{equation}

\noindent where $\rho_i$ is the density of the material composing the dust, $Q_{ext}$ is the
extinction efficiency (proportionality between geometric cross section and extinction
cross section), and $\omega$ is the dust albedo (ratio between scattering and extinction
cross section).  Both $Q_{ext}$ and $\omega$ depend on the optical constants of the dust
material and the ratio of the particle size to the wavelength.  For the ``astronomical
silicates'' and the graphite materials described by Draine \& Lee (1984) and for particle
sizes ranging between 0.1 and 1 $\mu$m, at near infrared 
wavelengths the product $\omega Q_{ext}$
varies from close to zero to a few.  We therefore expect a
scattering mass opacity to be of order $\kappa_s \sim 1/\rho_i a \sim 10^6 $cm$^2$ g$^{-1}$
for material density (in g cm$^{-3}$) and particle size (in microns) both of order unity.
The average mass density of dust grains in the nebula for a line of sight through the nebular 
lobes where $\tau \sim 1$ will then be of order 
$\bar{\rho}_d \sim 10^{-13}$ g cm$^{-3}$.

\par
\begin{figure}
  \plottwo{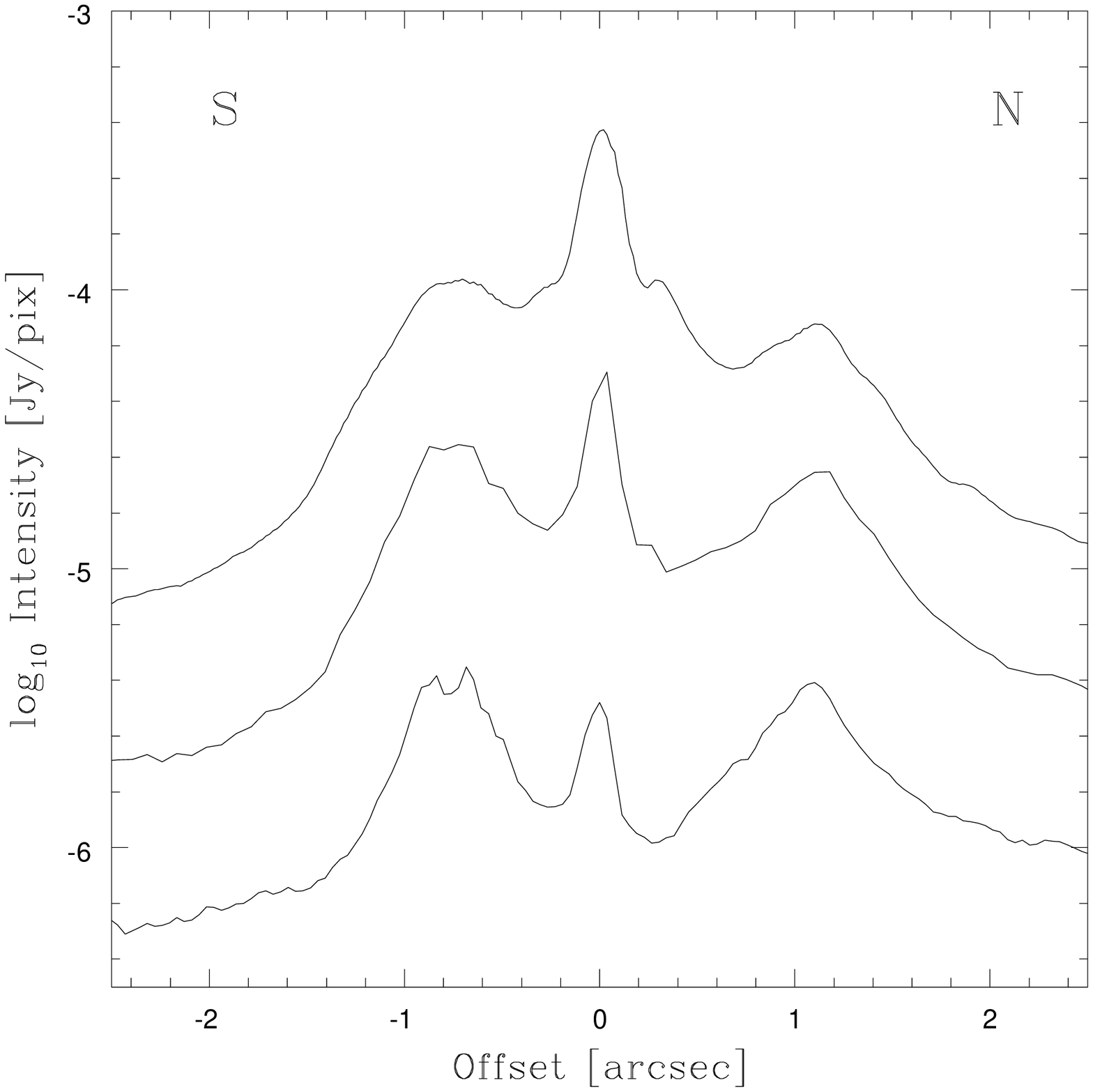}{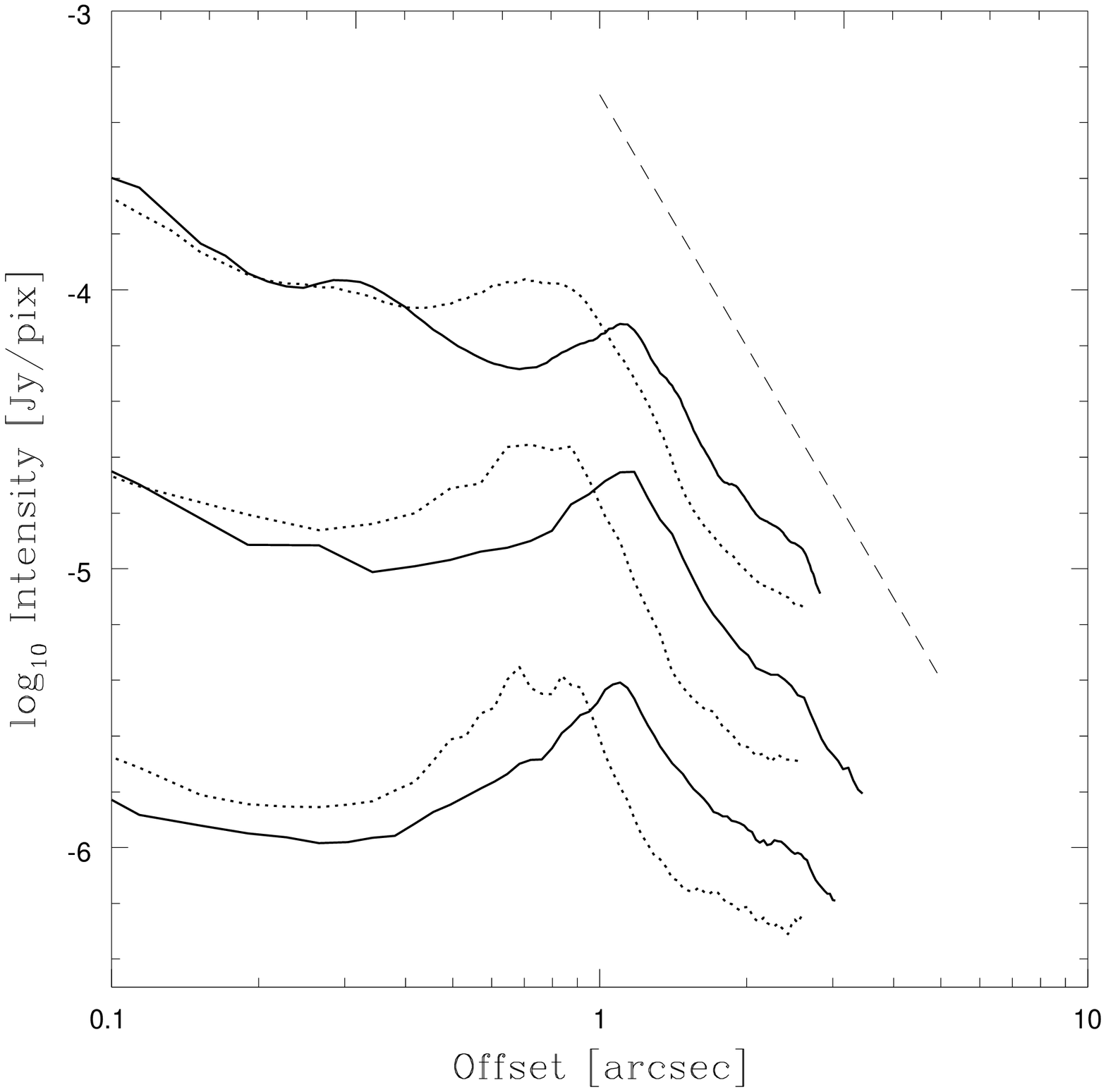}
  \caption[]{{\it Left:} Intensity is plotted along the nebular axis which is
    indicated in Figure \ref{f6} by the line {\em AB}. 
    The orientation is indicated by
    the letters N and S for north and south, respectively. From highest to lowest
    intensity is the F222M, F160W and F110W data. {\it Right:}
    The same intensity profiles are plotted on a log-log scale to emphasive
    the power law dependence of the surface brightness profile for distances from the
    central star of greater than  $\sim$1\arcsec.  The solid and dotted lines indicate
    the profiles for the north and the south lobes, respectively, while the dashed
    line shows a power law dependence $r^{\beta}$ with $\beta = -3.0$ for reference.
    \label{f10}}
\end{figure}

 In the optically thin scattering limit, the intensity variation along
the line of sight is proportional to the line of sight scattering optical
depth and the flux from the central illuminating source which is $F^* \propto L^* r^{-2}$,
modulo any internal extinction between the source and the scattering location.
The scattered light intensity variation with distance from the central source 
is then expected to vary as,

\begin{equation}
  \label{lobei}
  I \sim \bar{\rho}_d\Delta S r^{-2},
\end{equation}

\noindent where $r$ is the distance between the central source and
scattering location. The intensity variation along the nebular axis is plotted in
Figure \ref{f10} on a log-linear and log-log plot. The intensity is seen to 
drop off with distance from the central
source with a roughly power law dependence, $I \propto r^{\beta}$ with 
$\beta\approx$-3.0 for $r>1\arcsec$. Assuming that the lobe depth, 
$\Delta S$, is roughly constant in the scattering lobe then the 
inferred average line of sight dust density decreases inversely with
distance from the central source,
$\bar{\rho}_d \approx \bar{\rho}_{d,0}(r/r_0)^{-1}$.  

\par Integrating this distribution the {\em dust mass} in the scattering lobe is,
\begin{equation}
  M_l = \frac{\pi\Delta S^2 r_0}{4}\bar{\rho}_d(r_0) \ln(r_1/r_0) \sim 10^{27} \hbox{g},
\end{equation}

\noindent where the integration is taken between $r_0\sim 1\arcsec$ and $r_1 \sim 3\arcsec$,
a lobe width of $\Delta S \sim 2000$AU is used, and $\bar{\rho}_{d,0} = \bar{\rho}_d(r_0)$ is
taken as our previous order of magniture estimate, $\rho_d \sim 10^{-13}$ g cm$^{-3}$.
Inferring the total mass in the lobes requires knowledge of the gas to dust mass ratio, which may 
be much larger than the canonical value of $\sim$100 in the diffuse ISM due to grain destruction
processes that may be occurring in the outflow.

\par A great deal more information can be garnered from the infrared data presented
in this paper by comparing it to observations at other wavelengths.
For instance, high resolution CO observations of the 
nebula can provide a more robust map of the underlying 
mass distribution as well as the velocity structure. 
The velocity of the material can be used to test the hypothesis that we are
seeing an outflow nebula, as opposed to an illuminated evacuated cavity.
The  combination of a molecular map and near infrared scattered light images at 
comparable resolution should be able to provide strong constraints on the nature 
of the dust particles in the outflow by providing more robust estimates of
mass opacities. Dust properties may be varying strongly with location in the nebula
due to the interaction of the outflow with the ambient cloud material through shocks
and photodissociation processes.  These processes play an important role in the
life cycle of dust in the universe, and may be potentially useful probes of 
the outflow mechanism operating in YSOs.

\subsubsection{Nebular Inclination}

\par We conclude this section with a comment on the nebular inclination, noting that
asymmetries in the morphology between the north and south lobes can be used as a 
constraint. It is a general feature of bipolar reflection nebula
with mass distributions that have equatorial density enhancements that for small
inclinations (out of the plane of the sky) the lobe tilted towards the 
observer has (1) a higher surface brightness, (2) is broader, and (3) peaks closer to the 
central illuminating source.  These features arise from the relative
amounts of obscuration caused by the equatorial 
material in projection against the illuminated lobes.  All three of these features can 
be clearly seen in the contour plots presented in Figure \ref{f8} and the intensity profile 
in Figure \ref{f10} indicating that it is the {\it south lobe that is tilted towards the Earth}.

\subsection{Interaction with Environment}

Although the beam size in the $^{12}$CO J=(2-1) maps of Gomez et al (1992) 
is comparable to the size of the entire NICMOS field of view, interesting correlations in 
morphology exist between these two data sets. A finger of high intensity emission is discernible 
in the CO maps which is aligned with the WK34 bipolar nebulosity. This finger of 
CO emission sweeps off towards the east,
as though an extension of the similarly shaped infrared nebulosity seen in our images. 
The observed near infrared reflection nebulosity traces the polar axis of a 
fairly well collimated outflow emanating from WK34 on scales of $\sim 1000$ AU. The
ambient gas in the region may be redirecting the northern part of this 
outflow towards the east. This might explain why the geometric center of the large scale 
molecular outflow does not lie directly over the infrared cluster. The connection 
between the bipolar nebulosity and the larger scale outflow is further reinforced by 
the fact that the inclination of WK34 implied by the morphology of the
infrared nebulosity (S lobe tilted toward Earth) is consistent with the orientation of 
the large scale molecular outflow (S lobe is blue shifted).

\section{Summary \& Conclusions}

	We have presented high resolution polarimetric and broad band images of the 
infrared cluster in AFGL437 which reveals a well collimated bipolar nebula.  Our 
polarization measurements show that this nebula is being illuminated by predominantly 
one source, WK34. Photometry for this source is consistent with it being 
a {\em low luminosity protostar}. We note the following additional results.
(1) A dust model that contains too many large ($>1\mu$m) dust particles, such as the
KMH dust model which is successful in describing molecular cloud extinction, is
inconsistent with the levels of polarization that we observe in the reflection nebulosity
observed here.  This indicates that the population of dust grains is composed mainly of
smaller particles.  
(2) The equator of the observed bipolar nebula coincides with a ``polarization disk''
similar to that seen in other protostars and protoplanetary nebulae.  This feature
arises from the presence of an optically thick, nearly edge of torus of material.
(3) The observed bipolar nebula resides near the center of a much larger scale
bipolar CO outflow.  We find that the inclination and
alignment of the smaller scale, near-infrared bipolar nebula is similar to the
cluster scale CO outflow.
(4) The bipolar morphology, high aspect ratio (length to width), and high degrees of
linear polarization of the near infrared nebulosity that we have observed cannot be
reproduced with a collapsing envelope and evacuated cavity model which has been successful
in explaining the near infrared nebulae associated with many other embedded YSOs.

\par From these points we are led to a picture of AFGL437 in which a low luminosity
protostar is the source of a well collimated axisymmetric bipolar outflow of
gas and dust that is being illuminated by the central source.  The lobes
contain a low density population of small grains that scatter (thus polarizing)
light from the central star. The small dust grains in the outflow which differ from 
those typical of molecular clouds are likely a result of processes occurring near 
the protostar.

\epsscale{0.5}

\epsscale{1.0}
\epsscale{1.0}

\epsscale{0.8}
\epsscale{1.0}

\epsscale{0.8}
\epsscale{1.0}

\epsscale{0.4}
\epsscale{1.0}


\begin{deluxetable}{clcccc}
\tablecaption{Log of Observations}
\tablehead{
\colhead{Camera$^{a}$} & 
\colhead{Filter} & 
\colhead{$\Delta \lambda_{eff}$} & 
\colhead{Phot.Const.$^{b}$} &
\colhead{rms noise$^{c}$}&
\colhead{FWHM$^{c}$} \\
\colhead{-} & 
\colhead{-} & 
\colhead{$\mu m$} & 
\colhead{ $\mu$Jy / (ADU s$^{-1}$)} &
\colhead{ $\mu$Jy $pix^{-1}$}&
\colhead{arcsec}\\}
\startdata
NIC1 & F160W   & 0.4000 & 2.393566 & 0.04 &  0.12\\
\tableline
NIC2 & F110W   & 0.5920 &  1.823290 & 0.023 & 0.09\\
NIC2 & F160W   & 0.4030 &  2.070057 & 0.03  & 0.14\\ 
NIC2 & F222M   & 0.1432 &  5.280848 & 0.16  & 0.175\\
\tableline
NIC1 & POL0S   & 0.4750 &  6.995908 & 0.04 & 0.10\\
NIC1 & POL120S & ''     &  6.912439 & ''  & ''\\
NIC1 & POL240S & ''     &  6.914314 & ''  & ''\\ 
NIC2 & POL0L   & 0.2025 &  7.626095 & 0.2 & 0.175\\
NIC2 & POL120L & ''     &  7.529697 &  '' & ''\\
NIC2 & POL240L & ''     &  7.516947 &  '' & ''\\
\enddata
\tablenotetext{a}{The platescale for NIC1 is 0.043\arcsec pix$^{-1}$ and NIC2 is 
  0.076\arcsec pix$^{-1}$.}
\tablenotetext{b}{per pixel}
\tablenotetext{c}{The rms noise level was measured directly from a region in the 
  northern corner of the images. The FWHM is measured for the bright source, AFGL437N (see
  Figure \ref{f2} for source location).}

\end{deluxetable}

%
%

\begin{deluxetable}{lccc}
\tablecolumns{4}
\tablecaption{Aperture Polarimetry of North Lobe $^{a}$}
\tablehead{
\colhead{Filter} & 
\colhead{I} &
\colhead{P} &
\colhead{$\theta$} \\
\colhead{} &
\colhead{Jy} &
\colhead{\%} &
\colhead{deg.}\\}

\startdata
POL-S (1 $\mu m$) & 4.87 $\times 10^{-4}$  & 45 $\pm$ 3 & 24 $\pm$ 1 \\
POL-L (2 $\mu m$) & 7.0  $\times 10^{-3}$  & 65 $\pm$ 3 & 28 $\pm$ 1\\
\enddata
\tablenotetext{a}{A 0.5 \arcsec radius aperture centered on northen lobe of 
  bipolar nebulosity (see Figure 4).}

\end{deluxetable}

\begin{deluxetable}{llccc}
\tablenum{3}
\tablecolumns{5}
\tablecaption{Photometry of WK34 and Bipolar Lobes}
\tablehead{ 
  \colhead{} & 
  \colhead{Units} &
  \mcol{1}{c}{F222M NIC2} &   
  \mcol{1}{c}{F160W NIC2} &   
  \mcol{1}{c}{F110W NIC2}\\}

\startdata
WK34 &  Jy &

\mcol{1}{l}{2.30 $\pm 0.2 \times 10^{-3}$}&
\mcol{1}{l}{2.89 $\pm 0.3\times 10^{-4}$} & 
\mcol{1}{l}{5.54 $\pm 0.6 \times 10^{-6}$}\\

F$_{\nu}$ (North)& Jy pix$^{-1}$ &
\mcol{1}{l}{0.86\mult10$^{-4}$} &
\mcol{1}{l}{2.32\mult10$^{-5}$} &
\mcol{1}{l}{3.84\mult10$^{-6}$}\\

F$_{\nu}$ (South)\tablenotemark{a}& Jy pix$^{-1}$ &
\mcol{1}{l}{1.29\mult10$^{-4}$}&
\mcol{1}{l}{3.51\mult10$^{-5}$} &
\mcol{1}{l}{7.3752\mult10$^{-6}$}\\

\enddata

\tablecomments{The photometry for the north and south lobes is taken as the peak value
  in each lobe.}
\tablenotetext{a}{The peak surface brigthness in the southern lobe of the
  nebula in the F110W image is coincident with a very blue point like source which
  may be a foreground source.}
\end{deluxetable}


\begin{thebibliography}{bib1}
\bibitem[]{} Adams, F., Lada, C.J., Shu, F. 1987, ApJ, 312, 788
\bibitem[]{} Ageorges, N., \& Walsh, J.~R.\ 2000, \aap, 357, 661 
 \bibitem[]{} Arquilla, R., \& Goldsmith P.F. 1984, ApJ, 279, 664
\bibitem[]{} Bastien, P. \& Menard, F. 1988, ApJ, 326, 334
\bibitem[Cardelli, Clayton \& Mathis(1989)]{1989ApJ...345..245C} Cardelli, J.\ A., Clayton, G.\ C.\ \& Mathis, J.\ S.\ 1989, \apj, 345, 245 
\bibitem[]{} Cohen, M. \& Kuhim L.V. 1977, PASP, 89, 829
\bibitem[]{} Draine, B.T. \& Lee, H.M. 1984, ApJ, 285, 89
\bibitem[Goodman, Jones, Lada \& Myers(1995)]{1995ApJ...448..748G} Goodman, A.\ A., Jones, T.\ J., Lada, E.\ A.\ \& Myers, P.\ C.\ 1995, \apj, 448, 748 
\bibitem[]{} Gomez, J.F., Torelles, J.M., Estalalella, R., Anglada, G., Vades-Montenegro, L. and Ho, P.T.P. 1992, ApJ, 397, 492
\bibitem[]{} Hines, D.C., Schmidt, G.D., Schneider, G. 2000, PASP, 112, 983
\bibitem[]{} Kenyon, S.J., Whitney, B.A., Gomez, M. \& Hartmann, L. 1993, ApJ, 414, 773
\bibitem[]{} Kim, S., Martin, P.G. and Hendry, P.D. 1994, ApJ, 422, 164
\bibitem[]{} Mathis, Rumpl \& Nordsieck, K.H., 1977, ApJ, 217, 425 (MRN)
\bibitem[]{} Meakin, C.A., Bieging, J.H., Latter, W.B., Hora, J.L, Tielens, A.G.G.M. 2003, ApJ, 585, 482
\bibitem[]{} Pendleton, Y., Tielems, A.G.G.M, \& Werner, M.W. 1990, ApJ, 349, 107
\bibitem[]{} Sahai, R., Hines, D.C., Kastner, J.H., Weintraub, D.A., Trauger, J.T., Rieke, M.J., Thompson, R.I, \& Schneider, G. 1998, ApJL, 492, 163
\bibitem[Sparks \& Axon(1999)]{1999PASP..111.1298S} Sparks, W.\ B.\ \& Axon, D.\ J.\ 1999, \pasp, 111, 1298 
\bibitem[]{} Tereby, S., Shu, F., Cassen, P. 1984, ApJ, 286, 529
\bibitem[Thompson et al.(1998)]{1998ApJ...492L..95T} Thompson, R.\ I., Rieke, M., Schneider, G., Hines, D.\ C.\ \& Corbin, M.\ R.\ 1998, \apjl, 492, L95 
\bibitem[Thompson et al.(1999)]{1999AJ....117...17T} Thompson, R.\ I., Storrie-Lombardi, L.\ J., Weymann, R.\ J., Rieke, M.\ J., Schneider, G., Stobie, E.\ \& Lytle, D.\ 1999, \aj, 117, 17 
\bibitem[]{} Torelles,J.M., Gomez, J.F., Anglada, G., Estalallela, R., Mauersberger, R. and Eiroa, C. 1992, ApJ, 392, 616 


\bibitem[]{} Weintraub, D.\ A.\ \& Kastner, J.\ H.\ 1996a, ASP Conf.\ Ser.\ 97: Polarimetry of the Interstellar Medium, 345 
\bibitem[]{} Weintraub, D., Kastner, J.H., Gatley, I., Merill, K.M. 1996b, ApJL, 468, 45
\bibitem[]{} Weintraub, D., \& Kastner, J.H., 1996c, ApJ, 458, 670
\bibitem[]{} Weintraub, D.A., Kastner, J.H., Hines, D.C., \& Sahai, R. 2000a, ApJ, 531, 401
\bibitem[]{} Weintraub, D.\ A., Goodman, A.\ A.\ \& Akeson, R.\ L.\ 2000b, Protostars and Planets IV (Book - Tucson: University of Arizona Press; eds Mannings, V., Boss, A.P., Russell, S.\ S.), p.\ 247, 247 

\bibitem[Werner, Capps \& Dinerstein(1983)]{1983ApJ...265L..13W} Werner, M.\ W., Capps, R.\ W.\ \& Dinerstein, H.\ L.\ 1983, \apjl, 265, L13 
\bibitem[]{} Whitney, B. and Hartmann, L. 1993, ApJ, 402, 605
\bibitem[]{} Wood, K., Smith, D., Whitney, B., Stassun, K., Kenyon, S.J., Wolff, M.J., \& Bjorkman, K.S. 2001, ApJ, 561, 299
\bibitem[]{} Wynn-Williams, C.G., Becklin, E.E., Beichman, C.A., Capps, R. \& Shakeshaft, J.R. 1981, ApJ, 246, 801

\end{thebibliography}
\end{document}